\documentclass[pre,reprint,superscriptaddress,amsmath,amssymb,aps]{revtex4-2}

\usepackage{amsthm,amsfonts}
\usepackage{enumitem}
\usepackage{graphicx}
\allowdisplaybreaks

\usepackage{scalerel}
\usepackage{xcolor}
\usepackage{hyperref}
\usepackage[capitalise]{cleveref}
\AddToHook{cmd/appendix/before}{\crefalias{section}{appendix}}
\usepackage{libertinus}
\usepackage{libertinust1math}
\usepackage{bm}

\newcommand\bs[1]{\boldsymbol{#1}}
\newcommand{\wh}{\widehat}
\newcommand{\wt}{\widetilde}

\renewcommand{\P}{\mathbb{P}}
\newcommand{\E}{\mathbb{E}}
\DeclareMathOperator{\1}{\mathbf{1}}
\newcommand{\eps}{\varepsilon}

\DeclareMathOperator{\Var}{\mathrm{Var}}
\DeclareMathOperator{\Cov}{\mathrm{Cov}}
\DeclareMathOperator{\Corr}{\mathrm{Corr}}

\begin{document}


\title{Boltzmann-like occupation of nonequilibrium steady states on dense disordered networks}

\author{Jacob Calvert}
\altaffiliation{Present address: Department of Mathematical Sciences, University of Delaware, Newark, DE, USA}
\email{calvert@udel.edu}
\affiliation{Institute for Data Engineering and Science, Georgia Institute of Technology, Atlanta, GA, USA}
\affiliation{Santa Fe Institute, Santa Fe, NM, USA}
\affiliation{Department of Mathematics, London School of Economics, London, UK}

\begin{abstract}
A central problem in statistical physics is to extend the Boltzmann distribution to nonequilibrium steady states (NESS). We prove that NESS on large dense disordered networks have Boltzmann-like occupation despite extensive entropy production. One consequence is that such NESS can conceal their nonequilibrium nature from measurements of static response to energy perturbations. We further prove that the active-matter heuristic of ``low rattling'' is asymptotically exact for these NESS. Intuitively, they spend a greater fraction of their time in states they leave more slowly. We show that rattling effectively extends the Boltzmann distribution to such NESS, as well as the broader class of ``equiaccessible'' steady states, which play a role in our analysis akin to that of equilibrium in linear response.
\end{abstract}

\maketitle

\section{Introduction}

Usefully extending the Boltzmann distribution to nonequilibrium steady states (NESS) is a central problem of statistical physics \cite{Zia1995,Trepagnier2004,Derrida2007,PhysRevLett.100.030601,PhysRevE.85.041133,PhysRevE.92.052120}. The problem entails relating the steady-state probability of a microstate to measurable physical properties thereof \cite{Freitas2022}. A simple argument due to Landauer shows that no ``local'' property of states suffices to explain NESS occupation in general \cite{Landauer1975,Landauer1975stability,Landauer1988,MR3070751}. An alternative is to consider dynamical ensembles, or probability distributions on path space, that account for entropy flux and dynamical activity \cite{MR96415,MR106580,MR108922,PhysRevLett.100.030601,MR2605852,MR4080719}. However, this approach ultimately expresses steady-state probabilities in terms of sums over stochastic trajectories, which is impractical for computations \cite{Freitas2022}. One exception is the near-equilibrium regime, in which these sums reduce to Boltzmann factors modified by the transient part of the ``irreversible'' entropy flux, a result known as the McLennan formula \cite{MR2605852,MR2771859,MR4496392}. It is further known that systems in rapidly alternating contact with thermal baths at different temperatures, although technically out of equilibrium, have effective Boltzmann factors that are geometric means of those corresponding to each bath \cite{mamede2025exactmappingnonequilibriumequilibrium,10.1209/0295-5075/ae7ab7}.

Our main results establish that NESS on large dense disordered networks have occupation that is asymptotically indistinguishable from equilibrium despite extensive entropy production, and extend the Boltzmann distribution to these NESS in the sense of \cite{Freitas2022}. Our results imply that the steady-state distributions of such NESS can be inferred through local measurements made away from stationarity. They further establish that the active-matter heuristic of ``low rattling'' \cite{MR4269292}, as formulated in \cite{MR4826995}, is asymptotically exact for such NESS. We explain how the latter two properties are generic features of NESS whose states are nearly equally accessible, or {\em equiaccessible}, in a sense. We then use the Boltzmann-like steady-state distribution of NESS on dense disordered networks to estimate their entropy production (EP) and characterize their response to perturbations. The former shows that an approximate version of the minimal EP principle cannot hold for general NESS on dense networks, contradicting a recent claim \cite{Ray2025}. The latter leads to a nonequilibrium fluctuation--dissipation theorem (FDT), which implies that the static response of NESS on large dense disordered networks to energy perturbations is effectively indistinguishable from that of equilibrium systems \footnote{It is known that the response of general NESS to energy perturbations has the {\em form} of the static FDT \cite{PhysRevX.10.011066}. For NESS on large dense disordered networks, we show that even the {\em value} of the response to energy perturbations is asymptotically indistinguishable from that of the corresponding equilibrium. This result extends to more general perturbations of the transition rates.}.

\section{NESS on dense disordered networks}

We model the steady state as a Markov jump process on $N$ discrete states \cite{MR260355,MR443796,MR1966330,Zia_2007,PhysRevX.10.011066,MR4925514}. We express the transition rate from state $j$ to state $i(\neq j)$ as 
\begin{equation}\label{eq:rate x}
W_{ij} = e^{E_j} X_{ij},
\end{equation}
in terms of real-valued vertex parameters $E_j$ and edge weights $X_{ij} \in [0,\infty)$. For convenience, we set $X_{ii} = 0$. To make the column sums of $\bm{W} = (W_{ij})_{i, j}$ equal to zero, we define
\begin{equation}\label{eq:w def}
-W_{jj} = w_j = \sum_{i(\neq j)} W_{ij}.
\end{equation}
When $\bm{W}$ is irreducible, it has a unique stationary distribution $\bs{\pi}$, defined by $\bm{W} \bs{\pi} = \bm{0}$ and $\bm{1}^\top \bs{\pi} = 1$, which further satisfies $\pi_i > 0$ \cite{MR1600720}. In particular, when $\bm{X} = (X_{ij})_{i,j}$ is symmetric, $\bs{\pi}$ is the Boltzmann-like distribution
\begin{equation}\label{eq: boltzmann form}
\pi_i^{\mathrm{eq}} = \frac{e^{-E_i}}{Z},
\end{equation}
where the vertex parameters play the role of dimensionless energies in the partition function $Z = \sum_i e^{-E_i}$.

We study universal features of steady states on dense disordered networks by treating $\bm{X}$ in \cref{eq:rate x} as a random matrix \cite{doi:10.1137/1009001,MR972445,Timm2009RandomTransitionRate,MR2539237,MR3168123,PhysRevE.104.024305,NakerstDenisovHaque2023SparseGenerators,calvert2025localglobalcorrelationsdynamicsdisordered,xr67-xplb,Calvert2026}. Specifically, we fix arbitrary vertex parameters $(E_i)_i$ and assume that the edge-weight pairs $\{(X_{ij},X_{ji})\}_{i<j}$ 
are independent copies of a pair $(X,X')$ with the same distribution as $(X',X)$, and $\Var(X) \in (0,\infty)$. Equations~\eqref{eq:rate x}~and~\eqref{eq:w def} then define a random transition rate matrix $\bm{W}$. 

The parameterization in \cref{eq:rate x} is algebraically redundant for a fixed $\bm{W}$ (e.g., we can set all $E_i = 0$), but not for the ensemble of rate matrices we consider. The distribution of $\bm{W}$ uniquely determines the vertex parameters up to a global shift of $E_j \mapsto E_j - c$ and the distribution of edge weights up to a corresponding scaling $X_{ij} \mapsto e^c X_{ij}$ \footnote{{Suppose that $(\bm{E}, (X,X'))$ and $(\wt{\bm{E}}, (\wt{X},\wt{X}'))$ are two choices of vertex parameters and a representative edge-weight pair that satisfy the hypotheses and give rise, via \cref{eq:rate x}, to transition rate matrices $\bm{W}$ and $\wt{\bm{W}}$ with the same distribution. By equating the expected values of the entries of $\bm{W}$ and $\wt{\bm{W}}$, we find that $e^{E_j}\E[X] = e^{\wt{E}_j} \E[\wt{X}]$, hence $\wt{E}_j = E_j - c$ in terms of $c = \log (\E[\wt{X}]/\E[X])$, for every $j$. Since $W_{ij}$ and $\wt{W}_{ij}$ have the same distribution, the same is then true of $(\wt{X},\wt{X}')$ and $(e^c X, e^c X')$.}}. In particular, since $\bs{\pi}^{\mathrm{eq}}$ and the distribution of $\bs{\pi}$ are invariant under this transformation, the ensemble uniquely determines them.

We specify the distribution of $\bm{W}$ through the pair $(X,X')$ instead of individual edge weights so that we can enforce local detailed balance \cite{10.21468/SciPostPhysLectNotes.32}. However, this model is flexible enough to accommodate the opposite extreme of $\P(X>0,X'>0) = 0$, in which case {\em every} transition is irreversible. The assumption that a representative edge weight $X$ has finite variance ensures that the edge weights are not too heavy-tailed, which could otherwise effectively sparsen the underlying transition graph \cite{MR3656964}.

We have in mind the special case when edge parameters $B_{ij} = B_{ji}$ and $F_{ij} = -F_{ji}$ couple $X_{ij} = e^{-B_{ij}+F_{ij}/2}$ to $X_{ji}$. The assumptions of our model are satisfied, for example, when $(B_{ij})_{i<j}$ and $(F_{ij})_{i<j}$ are independent copies of mean-zero Gaussians $B$ and $F$, respectively \footnote{Specifically, $B$ and $F$ need to be nondegenerate random variables whose moment-generating functions $M_B(t) = \E[e^{tB}]$ and $M_F(u) = \E[e^{uF}]$ are finite for $t=-2$ and $u=1$. Indeed, the independence of $B$ and $F$ implies that
\begin{align}
    \E[X^2] &= \E \big[e^{-2(B-F/2)}\big] = M_B(-2) M_F(1).
\end{align}
The condition that $(X,X')$ has the same distribution as $(X',X)$ requires $F$ and $-F$ to have the same distribution.}. 
The corresponding transition rates have the form
\begin{equation}\label{eq:rate def}
W_{ij} = e^{E_j - B_{ij} + F_{ij}/2},
\end{equation}
which is a common model of stochastic thermodynamics \cite{MR1966330,MR4229016,PhysRevX.10.011066,MR4694249,MR5042990}. As Owen et al.\ explain, \cref{eq:rate def} is reminiscent of the Arrhenius law of reaction kinetics in an energy landscape with wells of depth $E_j$, separated by barriers of height $B_{ij}$, and driven by non-conservative forces $F_{ij}$ \cite{PhysRevX.10.011066}. In particular, $\bs{\pi} = \bs{\pi}^\mathrm{eq}$ when $\bm{F} = \bm{0}$.

The number of states $N$ implicitly indexes the edge weight matrix $\bm{X}$, which we view as the principal $N \times N$ submatrix of a single infinite array of edge weights satisfying the hypotheses. For each $N$, the state space of the NESS in our model has the connectivity of a random directed graph on $N$ vertices with edge set $\{(i,j): X_{ij} > 0\}$ \footnote{In the special case when $X$ and $X'$ are independent, this graph is a dense Erd\H{o}s--R\'enyi random directed graph, where each edge is independently present with probability $p = \P(X > 0) > 0$ \cite{MR2885424}.}. Since each pair of edge weights has the same distribution, a standard argument shows that this graph is almost surely strongly connected for all sufficiently large $N$ \cite{MR1633290}. For such $N$, the Markov jump process defined by \cref{eq:rate x} is irreducible and $\bs{\pi}$ is unique. We tacitly assume that $N$ is large enough for $\bs{\pi}$ to be unique for the rest of the paper. We mean all statements about the asymptotic behavior of $\bs{\pi}$ and other random quantities to hold in the same almost-sure sense, i.e., with a probability of~$1$ over the randomness of the infinite array of edge weights.

\section{Main results}

\begin{figure}
    \centering
    \includegraphics[width=\columnwidth]{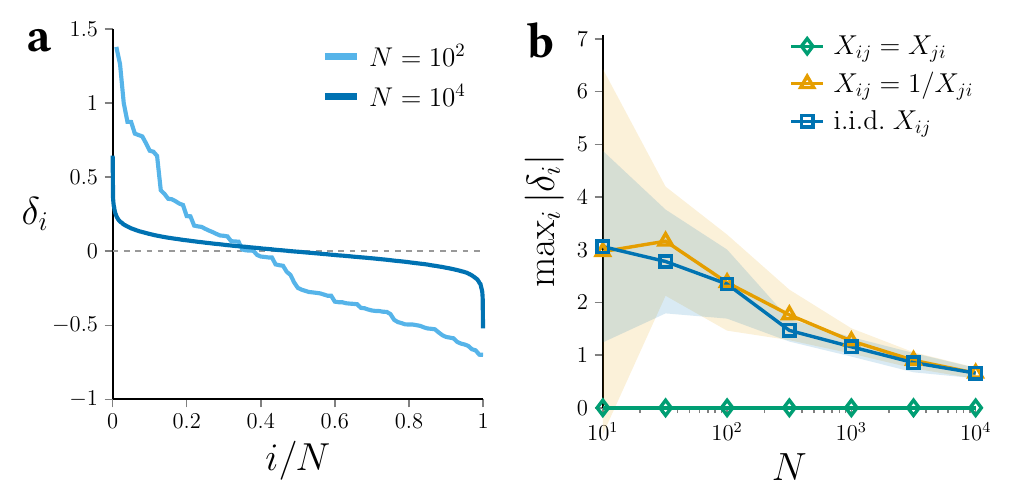}
    \caption{Boltzmann-like occupation of NESS on dense disordered networks. (a) Relative errors $\delta_i = \pi_i/\pi_i^\mathrm{eq} - 1$ in decreasing order when the $E_i$ and $\log X_{ij}$ are independent mean-zero Gaussians with respective variances of~$1$ and~$4$. (b) Maximum absolute relative error when the $\log X_{ij}$ are symmetric, anti-symmetric, or remain independent. The symmetric case produces transition rates in detailed balance, hence $\bs{\delta} = \bs{0}$. Markers and shading indicate means and $\pm 1$ standard deviation over $20$ trials.}
    \label{fig:1}
\end{figure}

Our first main result is that NESS on dense disordered networks have Boltzmann-like occupation probabilities when the state space is large enough. More precisely, $\bs{\pi}$ satisfies
\begin{equation}\label{eq:main result}
	\max_i \left| \frac{\pi_i}{\pi_i^\mathrm{eq}} - 1 \right| \to 0,
\end{equation}
as $N \to \infty$ (\cref{fig:1}). The convergence in \cref{eq:main result} is rather strong; it implies that the relative entropy and total variation distance between $\bs{\pi}$ and $\bs{\pi}^\mathrm{eq}$ tend to zero, along with all other R{\'e}nyi and $f$-divergences \cite{Polyanskiy_Wu_2025}. For convenience, we will write it in the equivalent form
\begin{equation}\label{eq: o1 result}
\pi_i = \pi_i^\mathrm{eq} (1+o(1)),
\end{equation}
in terms of a quantity $o(1)$ that tends to zero in absolute value, uniformly over $i$. When the edge weights further satisfy $\E[X^r] < \infty$ for some $r>2$, we can upgrade $o(1)$ to an explicit rate of $O(N^{-c})$, where $c$ is any number less than $\min\{\frac12,1-\frac{2}{r}\}$ (\cref{app: convergence rate}).

Equation~\eqref{eq:main result} is surprising because edge weights satisfying the assumptions can produce steady states that are far from equilibrium by various measures \cite{Zia_2007}. For example, recall that the EP associated with a distribution $\bm{p}$ on the states is defined as
\begin{equation}\label{eq: ep def}
    \sigma(\bm{p}) = \sum_{i \neq j} W_{ij} p_j \log \frac{W_{ij} p_j}{W_{ji} p_i}. 
\end{equation}
We will use \cref{eq:main result} to show that $\sigma_{\mathrm{NESS}} = \sigma(\bs{\pi})$ scales like $N^2/Z$ \footnote{To ensure that $\E[X\log (X/X')]$ is finite, we must further assume that $|\log X|$ is integrable.}. In particular, such NESS can exhibit increasingly Boltzmann-like occupation alongside extensive EP. This scenario is perfectly compatible with stochastic thermodynamics because it holds for edge weights that enforce local detailed balance, like those in \cref{eq:rate def}.

Before stating our second main result, we prove \cref{eq:main result} in the special case when $X$ and $X'$ are independent and have a finite third moment. We do so to demonstrate that it holds broadly for simple reasons. Under these further assumptions, \cref{eq:main result} is largely a consequence of a ``uniform'' strong law of large numbers (SLLN) applied to sums of edge weights and their products \cite[Lemma~2]{MR1235416}. Except for the last step, the proof of this special case is the same as the full proof, which we complete in \cref{app: full proof}.

\subsection{Proof of special case}

We build on a strategy introduced in \cite{Calvert2026}. According to basic Markov chain theory (e.g., \cite[Theorem 3.5.1]{MR1600720}), we can express $\bs{\pi}$ as
\begin{equation}\label{eq:pi pihat}
\pi_i \propto \frac{\wh{\pi}_i}{w_i},
\end{equation}
in terms of the exit rates $\bm{w}$ and the stationary distribution $\wh{\bs{\pi}}$ of the (discrete-time) Markov jump process with transition probabilities
\begin{equation}\label{eq:jump chain def}
\wh{W}_{ij} = \frac{W_{ij}}{w_j} = \frac{X_{ij}}{\sum_k X_{kj}}.
\end{equation}
Note that $\wh{\bm{W}}$ is irreducible because $\bm{W}$ is, hence $\wh{\bm{\pi}}$ exists and is unique. Equation~\eqref{eq:pi pihat} reflects the fact that the long-run fraction $\pi_i$ of time spent in state $i$ is proportional to the long-run fraction $\wh\pi_i$ of visits made there, times the typical duration $1/w_i$ of a visit \footnote{Interpreting $\protect{\wh{\bs{\pi}}}$ in this way requires $\protect{\wh{\bm{W}}}$ to be aperiodic.}.

Let $\mu = \E[X]$ denote the mean edge weight. The intuition behind \cref{eq:main result} is that each exit rate $w_i$ concentrates around its mean of $e^{E_i} \mu (N-1)$ because each state has many outgoing transitions, and $\wh{\bs{\pi}}$ is nearly uniform because the states are densely connected. More precisely, we will prove that $\bm{w}$ and $\wh{\bs{\pi}}$ satisfy
\begin{equation}\label{eq: two key estimates}
\frac{w_i}{e^{E_i} \mu N} = 1+o(1) \quad \text{and} \quad N \wh{\pi}_i = 1+o(1).
\end{equation}
Equation~\eqref{eq:pi pihat} then implies that $\pi_i$ equals
\begin{equation}\label{eq:proof idea}
\frac{\wh{\pi}_i / w_i}{\sum_k \wh{\pi}_k / w_k} = \frac{e^{-E_i} /\mu N^2}{\sum_k e^{-E_k}/\mu N^2} (1+o(1)) = \pi_i^\mathrm{eq}(1+o(1)),
\end{equation}
which proves \cref{eq:main result}.

We address the exit rates first. Denote the $j$-th column sum by $C_j = \sum_i X_{ij}$. Each $C_j$ is a sum of $(N-1)$ independent copies of $X$, which has a finite second moment. Under these conditions, the $N$-many column sums simultaneously converge to their expected values, due to a uniform SLLN \footnote{Note that only $\E[X]<\infty$ is necessary for $C_1 = \mu N (1+o(1))$ to hold almost surely. The finiteness of $\E[X^2]$ is needed to ensure that the $N$-many sums $(C_i)_i$ {\em simultaneously} converge. For a reference, see \cite[Lemma~2]{MR1235416}. Although the result is stated for an array of i.i.d.\ random variables, the proof in \cite{MR1235416} works as written in our case.}:
\begin{equation}\label{eq: row and col sum control}
C_i = \mu N(1+o(1)).
\end{equation}
The exit rates $w_i = e^{E_i} C_i$ consequently satisfy the first estimate in \cref{eq: two key estimates}.

We now specialize to the case when $X$ and $X'$ are independent and $\E[X^3] < \infty$ to establish the second estimate in \cref{eq: two key estimates}. (The full proof in \cref{app: full proof} continues from this point.) Under these further assumptions, the same SLLN ensures that all $N^2$-many sums of the form $\sum_j X_{ij} X_{jk}$ converge to their expected values as well:
\begin{equation}\label{eq: xx sums}
    \sum_j X_{ij} X_{jk} = \mu^2 N (1+o(1)).
\end{equation}
The entries of the two-jump transition matrix then satisfy
\begin{equation}\label{eq:intermediate w2}
(\wh{\bm{W}}^2)_{ik} = \sum_j \frac{X_{ij}}{C_j} \frac{X_{jk}}{C_k} = \frac{\sum_j X_{ij} X_{jk}}{\mu^2 N^2 (1+o(1))}  = \frac{1}{N} (1+o(1)).
\end{equation}
Since $\wh{\bs{\pi}}$ is also stationary for $\wh{\bm{W}}^2$, \cref{eq:intermediate w2} implies that
\begin{equation}
    \wh{\pi}_i = \sum_k (\wh{\bm{W}}^2)_{ik} \wh{\pi}_k = \frac{1}{N} (1+o(1)),
\end{equation}
which justifies the second estimate in \cref{eq: two key estimates}, completing the proof.

\subsection{Equiaccessible steady states}

Our second main result is that NESS on dense disordered networks satisfy
\begin{equation}\label{eq:pi nu}
\pi_i \propto \frac{1}{w_i} (1+o(1)) = e^{-\mathcal{R}_i} (1+o(1)),
\end{equation}
in terms of the quantity
\begin{equation}\label{eq: def of rattling}
\mathcal{R}_i = \log w_i.
\end{equation}
This follows from substituting the second estimate of \cref{eq: two key estimates} into \cref{eq:pi pihat}. Although the comparison of $\bs{\pi}$ to $\bs{\pi}^\mathrm{eq}$ in \cref{eq:main result} is perhaps more striking, \cref{eq:pi nu} is conceptually more important. It provides the simplest possible explanation of steady-state occupation: Such NESS spend a greater fraction of time in states they leave more slowly. As a consequence, the relative probabilities of states under $\bs{\pi}$ can be estimated from short trajectories initiated in the states of interest. In other words, $\mathcal{R}_i$ is a measurable property of a state that usefully extends the Boltzmann distribution to a broad class of NESS in the sense of \cite{Freitas2022}.

One interpretation of \cref{eq:pi pihat} is that the probability of a state is proportional to its accessibility times its stability, as respectively measured by $\wh\pi_i$ and $1/w_i$ \cite{MR3070751}. Equation~\eqref{eq:pi nu} and the simple explanation of occupation that it provides are then a common feature of NESS whose states are nearly equiaccessible in the sense that $\wh{\bs{\pi}}$ is uniform. Equiaccessible steady states correspond to doubly stochastic matrices because $\wh{\bs{\pi}} = \bm{u}$ if and only if the column-stochastic matrix $\wh{\bm{W}}$ is also row-stochastic \cite{MR4826995}. The parallel with equilibrium steady states, which correspond to reversible $\wh{\bm{W}}$ \cite{Zia_2007}, suggests expanding around the class of equiaccessible steady states as previous works on the near-equilibrium regime have done \cite{MR108922,MR2605852,MR2771859}. This is essentially what the full proof of \cref{eq:main result} does (see \cref{app: full proof}).

\section{Applications}

We proceed to demonstrate the conceptual value of our main results through several applications. In brief, they justify replacing $\bs{\pi}$ by a simpler distribution for a class of NESS beyond the near-equilibrium regime. Since these distributions are explicit while $\bs{\pi}$ is often not, doing so enables the evaluation of claims about NESS and makes new predictions as well.

\subsection{Principle of low rattling}

In their study of robot swarms, Chvykov et al.\ observed that plotting the logarithm of steady-state probabilities against a property of states called rattling produced a roughly linear scatter \cite{MR4269292}. They claimed that rattling would similarly anti-correlate with NESS occupation for many other systems, a principle they referred to as ``low rattling.'' As explained in \cite{MR4826995}, the quantity $\mathcal{R}_i$ in \cref{eq: def of rattling} is the analogue of rattling for Markov jump processes, and the observation of Chvykov et al.\ is analogous to high correlation between the {\em effective potential} $-\log \pi_{\bm{u}}$ and $\mathcal{R}_{\bm{u}}$, where the subscript $\bm{u}$ denotes a uniformly random state \footnote{Note that $\Corr(U,V)$ denotes the Pearson correlation coefficient, or normalized covariance, between $U$ and $V$.}:
\begin{equation}\label{eq:corr def}
\rho = \Corr (-\log \pi_{\bs{u}}, \mathcal{R}_{\bm{u}}) \approx 1.
\end{equation} 

Equation~\eqref{eq:corr def} can be viewed as an approximate analogue of the Boltzmann distribution because $\Corr(-\log \pi_{\bm{u}}^{\mathrm{eq}}, E_{\bm{u}}) = 1$ so long as the $E_i$ are not identical, in which case the correlation is undefined. In contrast, the correlation $\rho$ cannot be high for all steady states \cite{MR5019616}. It generally depends on the correlation between $-\log \wh{\pi}_{\bm{u}}$ and $\mathcal{R}_{\bm{u}}$, as well as the ratio of their variances \cite{MR4826995}. It is therefore important to identify broad classes of steady states for which $\rho \approx 1$.

\begin{figure}
    \centering
    \includegraphics[width=\columnwidth]{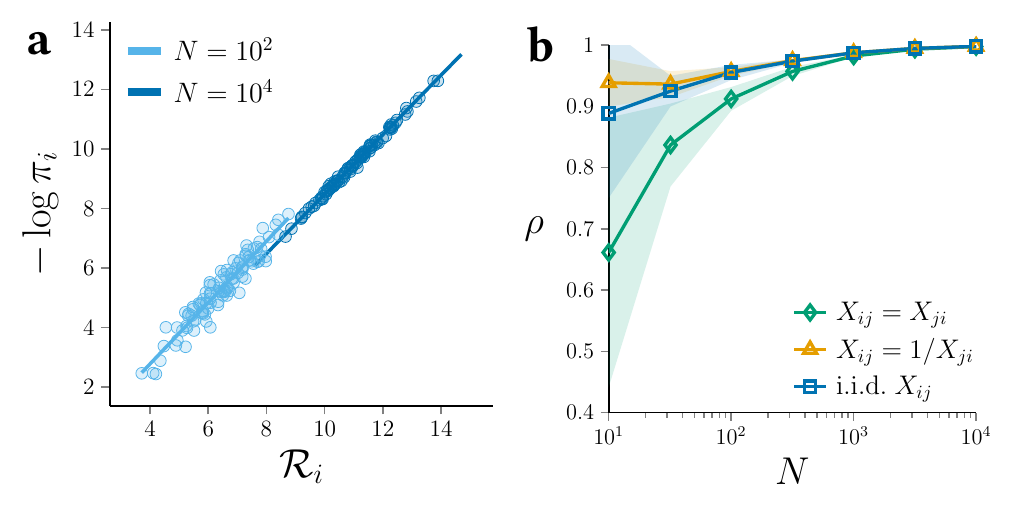}
    \caption{Rattling is asymptotically exact. (a) Scatter plots of effective potential versus rattling and (b) correlation coefficients for the conditions in \cref{fig:1}a and \cref{fig:1}b, respectively. Marks for $N=10^4$ in (a) are a uniformly random subset of $100$ states.}
    \label{fig:2}
\end{figure}

Under our model of a random NESS, $\rho$ is a random variable because the correlation in \cref{eq:corr def} is taken with respect to the uniformly random choice of state, conditioned on the random transition rate matrix $\bm{W}$. Equation~\eqref{eq:pi nu} implies that it satisfies
\begin{equation}\label{eq:corr result}
\rho = \Corr (\mathcal{R}_{\bm{u}} + o(1), \mathcal{R}_{\bm{u}}) \to 1,
\end{equation}
for any vertex parameters satisfying $\liminf_N \Var(E_{\bm{u}}) > 0$. In particular, the principle of low rattling as formulated in \cite{MR4826995} is asymptotically exact for NESS on dense disordered networks (\cref{fig:2}).

\subsection{Entropy production}

We now verify our earlier claim that $\sigma_{\mathrm{NESS}}$ scales like $N^2/Z$ and evaluate a recent conjecture about NESS on dense networks. To avoid degenerate cases, we assume that $a = \E[X\log(X/X')] \in (0,\infty)$, which requires $|\log X|$ to be integrable. We analyze $\sigma_{\mathrm{NESS}}$ via the EP associated with a {\em Gibbs tilt}, i.e., a probability distribution $\bm{p}^{(t)}$ of the form
\begin{equation}\label{eq:gibbs tilt def}
p^{(t)}_i = e^{-t E_i}/Z_t \quad\text{where}\quad Z_t = \sum_k e^{-t E_k}.
\end{equation}
For brevity, we denote the associated EP by $\sigma^{(t)} = \sigma (\bm{p}^{(t)})$.

By \cref{eq:main result}, $\bm{p}^{(1)}$ closely approximates $\bs{\pi}$, which suggests that $\sigma^{(1)}$ closely approximates $\sigma_{\mathrm{NESS}}$ \footnote{Although $\bs{\pi}^{\mathrm{eq}}$ and $\bm{p}^{(1)}$ are the same function of the vertex parameters, we intentionally use the latter in this section and in \cref{app: ep formulas} to avoid the confusion that might arise from writing $\sigma(\bs{\pi}^{\mathrm{eq}})$.}. A short calculation in \cref{app: ep formulas} confirms that $\sigma_{\mathrm{NESS}} = \sigma^{(1)} (1+o(1))$. To estimate $\sigma^{(1)}$, we recall the general expression for $\sigma(\bs{p})$ from \cref{eq: ep def} and observe that the probability flux $W_{ij} p_j^{(1)}$ equals $X_{ij}/Z$, because the factors involving the vertex parameters cancel. As a consequence, 
\begin{equation}\label{eq:ep of ess}
\sigma^{(1)} = \frac{1}{Z} \sum_{i < j} (X_{ij} - X_{ji}) \log \frac{X_{ij}}{X_{ji}}.
\end{equation}
The summands are independent, integrable, and have means of $2a$. By the standard SLLN, $\sigma^{(1)} = aN^2 (1+o(1))/Z$, hence
\begin{equation}\label{eq: ep of ness as}
\sigma_{\mathrm{NESS}} = \frac{aN^2}{Z} (1+o(1)).
\end{equation}

Next, we consider how $\sigma_{\mathrm{NESS}}$ compares to the minimal EP, defined as $\sigma_{\mathrm{min}} = \min_{\bm{p}} \sigma (\bm{p})$. Ray and Boyd simulated these quantities for a model closely related to \cref{eq:rate def} with random vertex and edge parameters \cite{Ray2025}. They observed that $\sigma_{\mathrm{NESS}}$ grew closer to $\sigma_{\mathrm{min}}$ as $N$ increased, and was only roughly $10\%$ larger than $\sigma_{\mathrm{min}}$ in their largest simulations. They conjectured that NESS on dense networks have nearly minimal EP when the number of states is large. To avoid some distracting technicalities, we will evaluate this claim by comparing the expected values of $\sigma_{\mathrm{NESS}}$ and $\sigma_{\mathrm{min}}$. However, the same conclusion will apply to the random quantities under further assumptions on $(X,X')$ and the vertex parameters.

We start with a formula for the expected EP associated with a Gibbs tilt. In \cref{app: ep formulas}, we show that
\begin{equation}\label{eq: mean ep for tilt}
    \E [ \sigma^{(t)} ] = \frac{NZ_{t-1}}{Z_t} (a (1-\tfrac{1}{N}) + \mu J_t),
\end{equation}
in terms of $J_t = (t-1) (\langle \bm{E} \rangle_{\bm{u}} - \langle \bm{E} \rangle_{\bm{p}^{(t-1)}})$, where $\langle \bm{E} \rangle_{\bm{p}} = \sum_i p_i E_i$. Since $\sigma_{\mathrm{min}} \leq \sigma^{(2)}$, \cref{eq: mean ep for tilt} implies that
\begin{equation}
    \E[\sigma_{\mathrm{min}}] \leq \frac{N Z_1}{Z_2} (a + \mu J_2).
\end{equation}
By Fatou's lemma \cite[Theorem~1.5.5]{MR3930614}, the almost-sure convergence in \cref{eq: ep of ness as} implies that $\E[\sigma_{\mathrm{NESS}}] \geq aN^2(1-o(1))/Z$ because $\sigma_{\mathrm{NESS}}$ is nonnegative. Hence, the ratio of expected values satisfies
\begin{equation}\label{eq: sigma12}
\frac{\E[\sigma_{\mathrm{NESS}}]}{\E[\sigma_{\mathrm{min}}]} \geq \frac{NZ_2}{Z_1^2} \frac{1}{1+ (\mu /a) J_2}(1-o(1)).
\end{equation}
To interpret this lower bound, note that 
\begin{equation}\label{eq:coeff of var}
\frac{N Z_2}{Z_1^2} = \frac{\E[b_{\bm{u}}^2]}{\E[b_{\bm{u}}]^2} \quad \text{and} \quad J_2 = \langle \bm{E} \rangle_{\bm{u}} - \langle \bm{E} \rangle_{\bm{p}^{(1)}},
\end{equation}
where $b_i = e^{-E_i}$. The first quantity is large when $\bm{p}^{(1)}$ concentrates on relatively few states, while the second measures the extent to which Boltzmann weighting decreases the mean of the vertex parameters.

\begin{figure}
    \centering
    \includegraphics[width=\columnwidth]{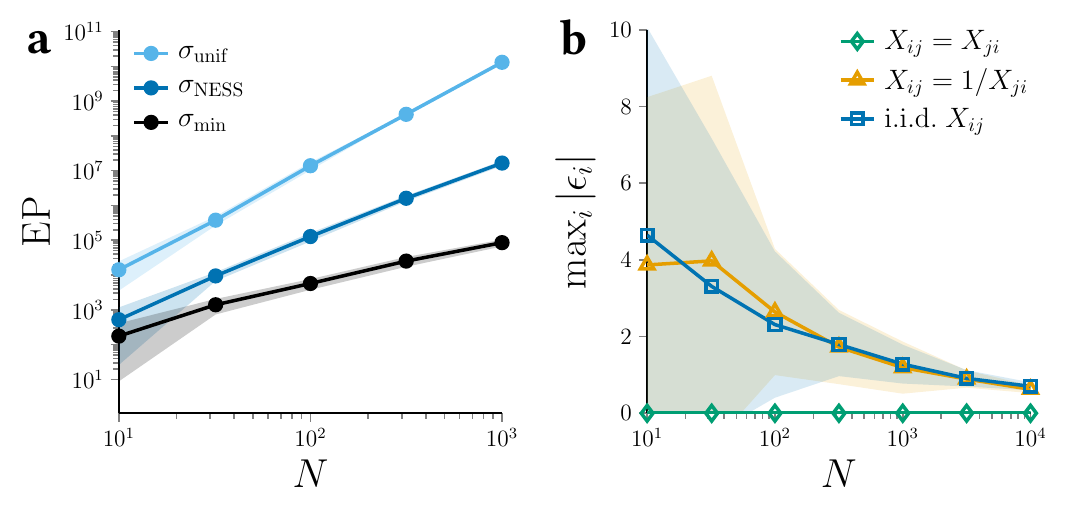}
    \caption{Consequences of \cref{eq:main result}. (a) The EP of the NESS grows faster with $N$ than the minimal EP, but more slowly than that of the uniform distribution, when $E_i = 2\log i$. (b) Vertex response is increasingly Boltzmann-like as measured by $\epsilon_i = \partial_{E_1} \pi_i / \partial_{E_1} \pi_i^{\mathrm{eq}} - 1$, under the same conditions as \cref{fig:1}b. Note that $\bs{\pi} = \bs{\pi}^{\mathrm{eq}}$ in the case of symmetric $X_{ij}$, hence $\bs{\epsilon} = \bs{0}$. Marks and shading indicate the mean and $\pm 1$ standard deviation over $20$ trials.}
    \label{fig:3}
\end{figure}

For example, if $E_i = 2\log i$, then $Z_1$ and $Z_2$ converge to positive constants, while $J_2 =O(\log N)$. In this case, \cref{eq: sigma12} implies that the expected value of $\sigma_{\mathrm{NESS}}$ is greater than that of $\sigma_{\mathrm{min}}$ by a factor of order $N/\log N$ (\cref{fig:3}a). This example demonstrates why a principle of near-minimal EP cannot generally hold for NESS on dense networks: The nearness of $\sigma_{\mathrm{NESS}}$ and $\sigma_{\mathrm{min}}$ is determined by the way the vertex parameters vary. 

\subsection{Nonequilibrium FDT}

As a further application of \cref{eq:main result}, consider the response of NESS occupation to perturbations of the vertex parameters. The steady-state distribution $\bs{\pi}$ of a Markov jump process with the transition rates in \cref{eq:rate x} is known to generally satisfy
\begin{equation}\label{eq:vertex perturb}
\partial_{E_i} \bs{\pi} = -\pi_i (\bm{e}_i - \bs{\pi}),
\end{equation}
where $\bm{e}_i$ is the $i$-th standard basis vector \cite{PhysRevX.10.011066,MR4694249}. Equation~\eqref{eq:main result} implies that NESS on dense disordered networks in particular satisfy
\begin{equation}\label{eq:vertex perturb1}
\partial_{E_i} \bs{\pi} = \partial_{E_i} \bs{\pi}^{\mathrm{eq}} (1+o(1)).
\end{equation}
Figure~\ref{fig:3}b shows that the relative error in \cref{eq:vertex perturb1} vanishes with $N$ analogously to that in \cref{fig:1}b.

Following \cite{PhysRevX.10.011066}, suppose the vertex parameters are $E_i = \beta \mathcal{E}_i(\eta)$, where $\beta = 1/k_B T$ in terms of Boltzmann's constant $k_B$, $T$ is the temperature, and $\mathcal{E}_i (\eta)$ is the system's underlying (free) energy landscape. Equation~\eqref{eq:vertex perturb1} implies that the response of an observable $\bm{O}$ to variation of $\eta$ is essentially unchanged by the presence of forces. Indeed, by the chain rule, the average of $\bm{O}$ with respect to $\bs{\pi}$ satisfies
\begin{equation}\label{eq:neq fdt pf1}
\partial_\eta \langle \bm{O} \rangle_{\bs{\pi}} = \sum_k O_k \partial_\eta \pi_k = \sum_k O_k \sum_i \partial_{E_i} \pi_k \, d_\eta E_i.
\end{equation}
Let $\bm{V} = -d_\eta \bs{\mathcal{E}}$ be the coordinate conjugate to $\eta$. Applying \cref{eq:vertex perturb1} to $\partial_{E_i} \pi_k$ and noting that $d_\eta E_i = -\beta V_i$, we find that the inner sum satisfies
\begin{equation}
    \sum_i \partial_{E_i} \pi_k \, d_\eta E_i = \beta \pi^{\mathrm{eq}}_k (V_k - \langle \bm{V} \rangle_{\mathrm{eq}})(1+o(1)),
\end{equation}
where $\langle \cdot \rangle_{\mathrm{eq}}$ abbreviates the equilibrium average $\langle \cdot \rangle_{\bs{\pi}^{\mathrm{eq}}}$. Substituting this into \cref{eq:neq fdt pf1} yields a nonequilibrium FDT:
\begin{align}
    \partial_\eta \langle \bm{O} \rangle_{\bs{\pi}} &= \beta \sum_k \pi^{\mathrm{eq}}_k O_k (V_k - \langle \bm{V} \rangle_{\mathrm{eq}})(1+o(1))\\ 
    &= \beta \Cov_{\mathrm{eq}}(\bm{O},\bm{V})(1+o(1)), \label{eq:neq fdt2}
\end{align}
where $\Cov_{\mathrm{eq}}$ denotes covariance with respect to $\boldsymbol{\pi}^{\mathrm{eq}}$. According to the standard (static) FDT, equilibrium steady states satisfy $\partial_\eta \langle \bm{O} \rangle_{\mathrm{eq}} = \beta \Cov_{\mathrm{eq}} (\bm{O},\bm{V})$. Equation~\eqref{eq:neq fdt2} therefore predicts that violations of the static FDT by NESS on dense disordered networks may not be detectable despite extensive EP.

\section{Conclusion}

We showed that, despite being far from equilibrium, NESS on large dense disordered networks have Boltzmann-like occupation (\cref{fig:1}). We proved this as a consequence of the exit rates $w_i$ concentrating around their expected values and the stationary probabilities $\wh{\pi}_i$ of the embedded process becoming increasingly uniform. The latter motivated the definition of equiaccessible steady states as those with uniform $\wh{\bs{\pi}}$, whose occupation is therefore determined by rattling $\mathcal{R}_i$. We then used these observations to evaluate existing claims and make new predictions about NESS. First, we established that the active-matter heuristic of low rattling is asymptotically exact for NESS on large dense disordered networks \cite{MR4269292,MR4826995} (\cref{fig:2}). Second, we argued that general NESS on dense networks cannot satisfy a principle of near-minimal EP, as was recently conjectured \cite{Ray2025} (\cref{fig:3}a). Third, we derived a nonequilibrium FDT, which implies that NESS on dense disordered networks can conceal their nonequilibrium nature from static response to energy perturbations.

Although we have emphasized what our results mean for steady states in general, one can use the ideas in this paper to establish many other interesting properties of NESS on dense disordered networks. For example, one can use the elements of the full proof of \cref{eq:main result} to show that the response of their occupation to more general perturbations of transition rates localizes. As a consequence, they asymptotically saturate the fundamental sensitivity bounds of Aslyamov and Esposito \cite[Eq.~16]{MR4694249}, which were previously known to be saturated only in the absence of current. One can further use \cref{eq:main result} to specialize the Hatano--Sasa equality \cite{PhysRevLett.86.3463} to versions of the nonequilibrium work relation and Clausius inequality \cite{PhysRevE.56.5018,PhysRevLett.78.2690,Crooks1998}, much in the same way that we obtained \cref{eq:vertex perturb1}. Trepagnier et al.\ raised the possibility of a nonequilibrium analogue of the Boltzmann distribution for precisely this reason \cite{Trepagnier2004}.

The virtue of our results about NESS on dense disordered networks, and equiaccessible steady states more broadly, is that they relate simple properties of states to global occupation. We focused on the ``forward'' direction of this relationship, which entails explaining occupation, as we did for the principle of low rattling. Future studies could explore the reverse direction, which entails designing or modifying dynamics to produce desirable occupation. This direction is inspired by the analogous use of the Boltzmann distribution to ``program'' the collective behavior of robots \cite{doi:10.1126/sciadv.abe8494} and microparticles \cite{yang_emergent_2022} through their individual properties, and parallels recent work on frenetic steering \cite{MR4579412,MR4755375}.

\section*{Acknowledgments}

We thank Dana Randall, James Holehouse, and David Wolpert for valuable discussions, and two anonymous referees for many helpful comments. We used Claude Sonnet 4.6 and Opus 4.8 (accessed May 2026) to optimize simulation scripts, and Claude Fable 5 (accessed June 2026) to proofread the manuscript before submission. All text is the author's. This work was supported by U.S.\ Army Research Office award MURI W911NF-19-1-0233 and National Science Foundation award CCF-2106687.

\appendix

\section{Properties of the transition graph}

A key input to the proof of \cref{eq:main result} in \cref{app: full proof} is the fact that there are many three-step paths of substantial weight between every pair of states. We make this precise as follows. Since a representative edge weight $X$ is nonnegative and has strictly positive variance, there exists $x > 0$ such that $p = \P(X \geq x) > 0$. We claim that, for all sufficiently large $N$, for every pair of states $(i,l)$, there are at least roughly $N^2$ three-step paths from $l$ to $i$ along which every edge has a weight of at least $x$:
\begin{equation}\label{eq:heavy path est}
\min_{i,l} \big| \{(j,k): X_{ij} \geq x, X_{jk} \geq x, X_{kl} \geq x\} \big| \geq \tfrac{1}{20} p^3 N^2.
\end{equation}
The constants in \cref{eq:heavy path est} are unimportant; we only need the lower bound to be comparable to $N^2$.

To prove \cref{eq:heavy path est}, we will use a version of Hoeffding's inequality \cite[Theorem~2.2.6]{MR3837109}. It states that if $Y_1, \dots, Y_n$ are independent and satisfy $Y_i \in [0,b]$, then
\begin{equation}\label{eq:hoeffding}
\P \left( \sum_{i=1}^n (Y_i - \E[Y_i]) \leq - t \right) \leq \exp (- 2t^2/b^2n).
\end{equation}
We start with a few applications. 

First, for all sufficiently large $N$, every state has at least $pN/2$ in-edges and at least $pN/2$ out-edges with weights of $x$ or more. The latter follows from a union bound over $j$ and an application of \cref{eq:hoeffding} with $t = p(N-1)/3$ to the independent indicators $Y_i = \1(X_{ij} \geq x)$, which have $\E[Y_i] = p$ and whose sum is the number of such out-edges of $j$:
\begin{equation}\label{eq:hoeffding app1}
\P \left( \min_j \sum_{i(\neq j)} Y_i \leq \tfrac{2}{3}p(N-1) \right) \leq N \exp (-\tfrac{2}{9}p^2(N-1)).
\end{equation}
Since the upper bound is summable over $N$, the Borel--Cantelli lemma implies that, for all sufficiently large $N$,
\begin{equation}\label{eq:hoeffding imp1}
\min_j \sum_{i(\neq j)} Y_i = \min_j \big| \{i: X_{ij} \geq x\} \big| \geq \tfrac23 p(N-1) \geq \tfrac12 pN.
\end{equation}
The same argument applies to the in-edges.

Second, consider for every pair of subsets $S, T \subseteq \{1,\dots,N\}$ the number of edges from $T$ to $S$ with weights of $x$ or more:
\begin{equation}
    e(S,T) = \big| \{(s,t) \in S \times T: X_{st} \geq x \} \big|.
\end{equation}
For all sufficiently large $N$, for every pair of subsets $|S|, |T| \geq pN/3$, we have that
\begin{equation}\label{eq:hoeffding app2}
e(S,T) \geq \tfrac{1}{5} p |S| |T|.
\end{equation}
To see why, fix such an $S$ and $T$, and group the terms of $e(S,T)$ according to the unordered pair of states they involve. The pair $\{u,v\}$ contributes
\begin{equation}\label{eq:unordered contrib}
\1(X_{uv} \geq x) \1 (u \in S, v \in T) + \1(X_{vu} \geq x) \1(v \in S, u \in T),
\end{equation}
which takes values in $[0,2]$. Because these contributions concern different (undirected) edges, they are independent. There are at most $\binom{N}{2}$ unordered pairs, and their sum has mean $p(|S||T| - |S \cap T|)$. Since $|S \cap T| \leq \sqrt{|S||T|}$, we have that
\begin{equation}\label{eq: e st mean}
\E [e(S,T)] - \tfrac{1}{5} p |S||T| \geq p \big( \tfrac45 |S||T| - \sqrt{|S||T|} \big) \geq \tfrac{3}{4}p|S||T|,
\end{equation}
for all sufficiently large $N$. By \cref{eq:hoeffding}, we find
\begin{equation}\label{eq: e st hoeffding}
\P \big( e(S,T) < \tfrac15 p |S||T| \big) \leq \exp \bigg( - \frac{2 (\tfrac{3}{4}p|S||T|)^2}{4 \binom{N}{2}} \bigg).
\end{equation}
Using the fact that $|S||T| \geq p^2 N^2/9$ and $\binom{N}{2} \leq N^2/2$, we can simplify the exponential rate of decay in \cref{eq: e st hoeffding} to $p^6 N^2/144$. There are fewer than $4^N$ pairs of sets to consider, and since $4^N \exp(- p^6 N^2/144)$ is summable, a union bound over $S, T$ and the Borel--Cantelli lemma together imply \cref{eq:hoeffding app2}.

We now return to proving \cref{eq:heavy path est}. Fix $i$ and $l$, and let
\begin{align}
S_{il} &= \{ j \notin \{i,l\} : X_{ij} \geq x \} \quad \text{and}\\
T_{il} &= \{k \notin \{i,l\}: X_{kl} \geq x\}.
\end{align}
By \cref{eq:hoeffding imp1}, both sets have at least $pN/2$ elements, so \cref{eq:hoeffding app2} applies to them. Every pair $(j,k)$ counted by $e(S_{il},T_{il})$ satisfies $j \neq k$ and $j,k \notin \{i,l\}$, and all edges along $l \to k \to j \to i$ have weights of at least $x$. The count of such paths is therefore at least
\begin{equation}\label{eq: path count 2}
e(S_{il},T_{il}) \geq \tfrac15 p (\tfrac12 p N)^2 =\tfrac{1}{20} p^3 N^2,
\end{equation}
for all sufficiently large $N$. Since this argument simultaneously applies to all pairs $(i,l)$, \cref{eq: path count 2} implies \cref{eq:heavy path est}, completing the proof.

\section{Proof of \texorpdfstring{\cref{eq:main result}}{Eq.~5}}\label{app: full proof}

The preceding proof of \cref{eq:main result} made additional assumptions to prove that
\begin{equation}\label{eq: remaining estimate}
\max_i |N \wh\pi_i -1| \to 0,
\end{equation}
as $N \to \infty$. The proof was otherwise complete. We now establish \cref{eq: remaining estimate} in full---that is, assuming only that the $(X_{ij},X_{ji})$ are independent copies of $(X,X')$, which has the same distribution as $(X',X)$, and that $\Var(X) \in (0,\infty)$.

The key idea is to analyze the three-jump transition probability matrix $\bm{P} = \wh{\bm{W}}^{\,3}$. We will show that there is a constant $\alpha \in (0,1]$ such that, for all sufficiently large $N$, $\bm{P}$ satisfies
\begin{equation}\label{eq:full proof prop1}
\min_{i,j} P_{ij} \geq \frac{\alpha}{N},
\end{equation}
and that the row sums of $\bm{P}$ satisfy
\begin{equation}\label{eq:full proof prop2}
\delta = \max_i \left| \sum_j P_{ij} - 1 \right| \to 0.
\end{equation}
The following argument explains why these two facts suffice to prove \cref{eq: remaining estimate}.

Due to \cref{eq:full proof prop1}, we can express $\bm{P}$ in terms of the uniform matrix $\bm{U} = \bm{1}\bm{1}^\top/N$ and a second column-stochastic matrix $\bm{K}$ as
\begin{equation}\label{eq:puk}
\bm{P} = \alpha\bm{U} + (1-\alpha) \bm{K}.
\end{equation}
Using the fact that $\wh{\bs{\pi}}$ is also stationary for $\bm{P}$, as well as the fact that $\bm{U} \bm{p} = \bm{u}$ for any probability (column) vector $\bm{p}$, it is easy to verify the identity
\begin{equation}\label{eq:col vec identity}
(\bm{I} - (1-\alpha) \bm{K} ) (\wh{\bs{\pi}} - \bm{u}) = \bm{P} \bm{u} - \bm{u}.
\end{equation}
The spectral radius of $(1-\alpha) \bm{K}$ is bounded above by its maximum row sum, which by \cref{eq:full proof prop2} is at most $1-\alpha+\delta$. Since $\delta < \alpha$ for all sufficiently large $N$, the Neumann series $\sum_{n=0}^\infty ((1-\alpha)\bm{K})^n$ converges for such $N$, and multiplying it on the left of \cref{eq:col vec identity} yields
\begin{equation}\label{eq:inverted k}
\wh{\bs{\pi}} - \bm{u} = \left(\sum_{n = 0}^\infty ((1-\alpha) \bm{K})^n \right) (\bm{P} \bm{u} - \bm{u}).
\end{equation}
For any $N \times N$ nonnegative matrix $\bm{M}$ and $N \times 1$ column vector $\bm{v}$, we have the bound
\begin{equation}\label{eq: general norm bd}
\| \bm{M} \bm{v} \|_\infty \leq \|\bm{v}\|_\infty \max_i \sum_j M_{ij},
\end{equation}
where $\|\bm{v}\|_\infty = \max_i |v_i|$. In particular, 
\begin{equation}\label{eq: norm bd}
    \|(1-\alpha)\bm{K} \bm{v}\|_\infty \leq \|\bm{v}\|_\infty (1-\alpha + \delta). 
\end{equation}
 By iteratively applying this bound to \cref{eq:inverted k}, we find that
\begin{equation}\label{eq: pi u inf bd1}
\|\wh{\bs{\pi}} - \bm{u}\|_\infty \leq \|\bm{P} \bm{u} - \bm{u}\|_\infty \sum_{n=0}^\infty (1-\alpha + \delta)^n \leq  \frac{\|\bm{P} \bm{u} - \bm{u}\|_\infty}{\alpha - \delta}.
\end{equation}
Equation~\eqref{eq: remaining estimate} then follows from \cref{eq:full proof prop2}: 
\begin{equation}\label{eq: pi u inf bd2}
N \|\wh{\bs{\pi}} - \bm{u}\|_\infty \leq \frac{N\|\bm{P} \bm{u} - \bm{u}\|_\infty}{\alpha - \delta} \leq \frac{\delta}{\alpha-\delta} \to 0.
\end{equation}
To complete the proof, it remains to explain why $\bm{P}$ satisfies \cref{eq:full proof prop1,eq:full proof prop2}.

First, the column sum estimate in \cref{eq: row and col sum control} implies that
\begin{equation}\label{eq: p3 lb1}
P_{il} = \sum_{j,k} \frac{X_{ij}}{C_j} \frac{X_{jk}}{C_k} \frac{X_{kl}}{C_l} \geq \frac{\sum_{j,k} X_{ij} X_{jk} X_{kl}}{\mu^3 N^3 (1+o(1))}.
\end{equation}
By \cref{eq:heavy path est}, for all sufficiently large $N$, there are at least $p^3 N^2/20$ directed paths of the form $l \to k \to j \to i$ along which every edge has a weight of at least $x$. In particular, the sum on the right-hand side of \cref{eq: p3 lb1} satisfies
\begin{equation}\label{eq: p3 lb2}
\sum_{j,k} X_{ij} X_{jk} X_{kl} \geq \tfrac{1}{20}p^3 x^3 N^2,
\end{equation}
hence \cref{eq:full proof prop1} holds with, e.g., $\alpha = p^3x^3/40\mu^3$.

Second, note that the argument leading to \cref{eq: row and col sum control} also shows that the row sums $R_i = \sum_j X_{ij}$ of $\bm{X}$ uniformly satisfy
\begin{equation}\label{eq: row sum control}
R_i = \mu N (1+o(1)).
\end{equation}
The estimates of $C_i$ and $R_i$ together imply that
\begin{equation}\label{eq:intermediate w1}
\sum_j \wh{W}_{ij} = \sum_j \frac{X_{ij}}{C_j} = \frac{R_i}{\mu N (1+o(1))} = 1+o(1).
\end{equation}
In other words, the vector $\bs{\varepsilon} = \wh{\bm{W}} \bm{1} - \bm{1}$ of row sum excesses has $\|\bs{\varepsilon}\|_\infty = o(1)$. We apply this fact to
\begin{equation}\label{eq:row sum exp}
\bm{P}\bm{1}-\bm{1} = \bs{\varepsilon} + \wh{\bm{W}} \bs{\varepsilon} + \wh{\bm{W}}^2 \bs{\varepsilon}.
\end{equation}
Combining \cref{eq: general norm bd,eq:row sum exp}, we find that
\begin{equation}\label{eq:delta est}
\delta = \|\bm{P}\bm{1}-\bm{1}\|_\infty \leq \left(1+ (1+\|\bs{\varepsilon}\|_\infty) + (1+\|\bs{\varepsilon}\|_\infty)^2 \right) \|\bs{\varepsilon}\|_\infty.
\end{equation}
This establishes \cref{eq:full proof prop2} because $\|\bs{\varepsilon}\|_\infty = o(1)$, completing the proof of \cref{eq:main result}. \qed

We briefly mention that the rationale for analyzing $\bm{P} = \wh{\bm{W}}^{\,3}$ is that neither $\wh{\bm{W}}$ nor $\wh{\bm{W}}^{2}$ satisfies a lower bound like \cref{eq:full proof prop1}. It cannot hold for $\wh{\bm{W}}$ because $\wh{W}_{ii} = 0$ by definition, while $(\wh{\bm{W}}^{2})_{ii} = 0$ when $(X,X')$ is such that $\P(X>0,X'>0) = 0$. Barring this extreme coupling of edge weights, an analogous argument with $\bm{P} = \wh{\bm{W}}^{2}$ works.

\section{An explicit convergence rate}\label{app: convergence rate}

We now upgrade the convergence in \cref{eq:main result} under the further assumption that $\E[X^r] < \infty$ for some $r > 2$. To be precise, letting $c < \min\{\frac{1}{2},1 - \frac{2}{r}\}$, we will prove that
\begin{equation}\label{eq:main result with rate}
\max_i \left| \frac{\pi_i}{\pi_i^{\mathrm{eq}}} - 1 \right| = O(N^{-c}).
\end{equation}

We begin by reducing \cref{eq:main result with rate} to an estimate of the column sums of $\bm{X}$:
\begin{equation}\label{eq: rate step1}
    \max_j \left| C_j - \mu N \right| = O(N^{1-c}).
\end{equation}
Equation~\eqref{eq: rate step1} improves the $o(1)$ term in the first estimate of \cref{eq: two key estimates} to $O(N^{-c})$. If we could similarly improve the second key estimate to
\begin{equation}\label{eq: rate step2}
    \max_i | N \wh{\pi}_i - 1 | = O(N^{-c}),
\end{equation}
then \cref{eq:proof idea} would imply \cref{eq:main result with rate}. In fact, this estimate also follows from \cref{eq: rate step1}, as we now explain. First, recalling the roles of $\alpha$, $\delta$, and $\bs{\eps}$ from \cref{app: full proof}, we have that
\begin{equation}\label{eq: rate step3}
    \max_i | N \wh{\pi}_i - 1 | \leq \frac{\delta}{\alpha - \delta} = O(\delta) = O(\|\bs{\eps}\|_\infty).
\end{equation}
These bounds are respectively justified by \cref{eq: pi u inf bd2}; by \cref{eq:intermediate w1,eq:delta est}, which imply that $\delta$ is less than $\alpha/2$ for all sufficiently large $N$; and by \cref{eq:delta est}. Second, since $\bm{X}$ and $\bm{X}^\top$ have the same distribution, \cref{eq: rate step1} implies the same estimate for the row sums of $\bm{X}$. Third, we revisit \cref{eq:intermediate w1}, finding that
\begin{equation}
    \sum_j \wh{W}_{ij} = \sum_j \frac{X_{ij}}{C_j} = \frac{R_i}{\mu N (1+O(N^{-c}))} = 1 + O(N^{-c}). 
\end{equation}
In particular, $\|\bs{\eps}\|_\infty = O(N^{-c})$, which by \cref{eq: rate step3} implies \cref{eq: rate step2}. 

It remains to establish \cref{eq: rate step1}. We will bound the deviation of the $j$-th column sum from its expected value as
\begin{equation}\label{eq: three term bd}
    \big| C_j - \mu N \big| \leq \big| C_j - \wt{C}_j \big| + \big| \wt{C}_j - \E[\wt{C}_j] \big| + \big| \E[\wt{C}_j] - \mu N \big|,
\end{equation}
where $\wt{C}_j = \sum_i \wt{X}_{ij}$ is the $j$-th column sum of $\wt{\bm{X}}$, the matrix of truncated edge weights $\wt{X}_{ij} = X_{ij} \1 (X_{ij} \leq N^{2/r})$. By a standard argument, the stronger moment condition of $\E[X^r] < \infty$ implies that $\max_{i,j \leq N} X_{ij} = o(N^{2/r})$ \footnote{Enumerate the off-diagonal entries of $\bm{X}$ so that the $n(n-1)$ entries with $i,j \leq n$ appear first, and call the resulting sequence $(Y_k)_{k \geq 1}$. By the Borel--Cantelli lemma, it suffices for $\P(Y_k > \lambda k^{1/r})$ to be summable over $k$ for every $\lambda > 0$, as then $\max_{i,j \leq N} X_{ij} = \max_{k \leq N(N-1)} Y_k = o(N^{2/r})$ almost surely. Each $Y_k$ is equal in distribution to $X$, so this follows from the equivalence of $\E[X^r] < \infty$ and the finiteness of $\sum_k \P(X > \lambda k^{1/r})$ for every $\lambda > 0$.}, hence $|C_j - \wt{C}_j|$ is eventually zero and can be ignored. 
Concerning the final term in \cref{eq: three term bd}, note that $(XN^{-2/r})^{r-1} > 1$ when $X > N^{2/r}$, hence
\begin{equation}\label{eq:xtil bd}
    \big| \wt{\mu} - \mu \big| = \E \big[ X \1 (X > N^{2/r}) \big] \leq N^{-2(r-1)/r} \, \E[X^r],
\end{equation}
where $\wt{\mu}$ denotes the expected value of the truncated edge weights. Since $2(r-1)/r > 1 > c$, \cref{eq:xtil bd} implies that
\begin{equation}
    \big| \E[\wt{C}_j] - \mu N \big| \leq N \big| \wt{\mu} - \mu \big| + \mu = o(N^{1-c}).
\end{equation}
We will use Bernstein's inequality to bound the remaining term in \cref{eq: three term bd}.

Bernstein's inequality states that, if $Z_1, \dots, Z_N$ are independent mean-zero random variables that are absolutely bounded as $|Z_i| \leq b$ for all $i$, then
\begin{equation}\label{eq:bernstein}
    \P \Big( \Big| \sum_i Z_i \Big| > t \Big) \leq 2 \exp \left( - \frac{t^2}{2(v + bt/3)} \right),
\end{equation}
for every $t \geq 0$, where $v = \sum_j \E[Z_j^2]$ \cite[Theorem~2.8.4]{MR3837109}. We fix a column $j$ and apply Bernstein's inequality to the random variables $Z_i = \wt{X}_{ij} - \wt{\mu}$ for $i \neq j$. Note that the $Z_i$ are absolutely bounded by $b = N^{2/r}$ and their total variance is $v = (N-1) \Var (\wt{X}_{ij}) \leq N \E[X^2]$. Choosing $t = N^{1-c}$, the exponential rate of decay in \cref{eq:bernstein} satisfies
\[
    \frac{t^2}{2(v + bt/3)} \geq \frac{N^{2(1-c)}}{2 (N\E[X^2] + N^{2/r + 1 - c}/3)}.
\]
The power of $N$ in the numerator is greater than in the denominator, because $c < \min\{\frac12, 1 - \frac{2}{r}\}$ implies $2(1-c) > \max\{1, \frac{2}{r} + 1 - c\}$. Consequently, there exist positive numbers $\kappa$ and $\gamma$ such that
\begin{equation}
    \P \big( \big| \, \wt{C}_j - \E[\wt{C}_j] \,\big| > N^{1-c} \big) \leq 2 \exp(-\kappa N^\gamma).
\end{equation}
By a union bound over $j$, the probability that some $\wt{C}_j$ deviates from its expected value by more than $N^{1-c}$ is at most $2N e^{-\kappa N^\gamma}$. Since this upper bound is summable over $N$, the Borel--Cantelli lemma implies that
\begin{equation}
    \max_j \big| \, \wt{C}_j - \E[\wt{C}_j] \, \big| = O(N^{1-c}),
\end{equation}
almost surely. Following the preceding discussion, this establishes \cref{eq: rate step1}, completing the proof.

\section{Entropy production formulas}\label{app: ep formulas}

Recall from \cref{eq:gibbs tilt def} that $\bm{p}^{(t)}$ is the probability distribution with $p_i^{(t)} \propto e^{-tE_i}$ and that $\sigma^{(t)} = \sigma (\bm{p}^{(t)})$ denotes the associated EP. First, we prove that $\sigma_{\mathrm{NESS}} = \sigma^{(1)} (1+o(1))$, which we used to obtain \cref{eq: ep of ness as}. We start by expanding the logarithm in $\sigma_{\mathrm{NESS}}$:
\begin{equation}
    \sigma_{\mathrm{NESS}} = \sum_{i \neq j} W_{ij} \pi_j \left( \log \frac{X_{ij}}{X_{ji}} + (E_j - E_i) + \log \frac{\pi_j}{\pi_i} \right).
\end{equation}
Note that, for any real-valued function $\bm{f}$ on the states, the stationarity of $\bs{\pi}$ implies that
\begin{equation}
    \sum_{i \neq j} W_{ij} \pi_j (f_j - f_i) = 0.
\end{equation}
We apply this fact with $f_j = E_j$ and $f_j = \log \pi_j$ to find that
\begin{equation}
    \sigma_{\mathrm{NESS}} = \sum_{i \neq j} W_{ij} \pi_j \log \frac{X_{ij}}{X_{ji}}.
\end{equation}
Note that the probability fluxes under $\bs{\pi}$ and $\bm{p}^{(1)}$ differ by $W_{ij} (\pi_j - p_j^{(1)}) = X_{ij} \delta_j/Z$, where $\delta_j = \pi_j/p_j^{(1)} - 1$ is the relative error in \cref{eq:main result}. The absolute difference in their EPs is then
\begin{equation}
    |\sigma_{\mathrm{NESS}} - \sigma^{(1)} | \leq \|\bs{\delta}\|_\infty \frac{1}{Z} \sum_{i \neq j} X_{ij} \, | \log (X_{ij}/X_{ji}) |.
\end{equation}
As in \cref{eq:ep of ess}, the standard SLLN implies that the sum on the right-hand side equals $\tilde{a}N^2 (1+o(1))/Z$, where $\tilde a = \E[X \cdot |\log (X/X')|] \in (0,\infty)$. Consequently, $\sigma_{\mathrm{NESS}}$ satisfies
\begin{equation}
    \sigma_{\mathrm{NESS}} = \sigma^{(1)} \left(1 + \|\bs{\delta}\|_\infty \frac{\tilde{a}}{a} (1+o(1)) \right),
\end{equation}
which verifies the claim because $\|\bs{\delta}\|_\infty = o(1)$ by \cref{eq:main result}.

We now derive \cref{eq: mean ep for tilt}. We start by substituting $W_{ij} = e^{E_j} X_{ij}$ into the definition of the EP from \cref{eq: ep def}. Expanding the logarithm, we find that
\begin{equation}
	\sigma(\bm{p}) = \sum_{i \neq j} X_{ij} r_j \log \frac{X_{ij}}{X_{ji}} + \sum_{i \neq j} X_{ij} r_j \log \frac{r_j}{r_i},
\end{equation}
where $r_i = p_i e^{E_i}$. The expected value of $\sigma(\bm{p})$ with respect to the random edge weights therefore satisfies
\begin{align}
	\E[\sigma(\bm{p})] &= a \sum_{i \neq j} r_j + \mu \sum_{i \neq j} r_j \log \frac{r_j}{r_i}\\ &= a (N-1) S(\bm{q}) + \mu S(\bm{q}) \sum_{i \neq j} q_j \log \frac{q_j}{q_i},
\end{align}
in terms of the related probability distribution $q_i = r_i/S(\bm{q})$ with normalization constant $S (\bm{q}) = \sum_k r_k$. Some algebra shows that $\sum_{i \neq j} q_j \log (q_j/q_i)$ equals $N J(\bm{q}\|\bm{u})$, where $J(\bm{q} \| \bm{u}) = D(\bm{q} \| \bm{u}) + D(\bm{u} \| \bm{q})$ is the sum of Kullback--Leibler divergences between $\bm{q}$ and the uniform distribution, hence 
\begin{equation}\label{eq: mean ep em}
	\E[\sigma(\bm{p})] =  NS(\bm{q}) \left( a(1-\tfrac{1}{N}) + \mu J(\bm{q}\|\bm{u}) \right).
\end{equation}

In the special case when $\bm{p}$ has entries $p^{(t)}_i = e^{-t E_i}/Z_t$ with normalization constant $Z_t = \sum_k e^{-t E_k}$, the corresponding $\bm{q}^{(t)} = \bm{p}^{(t-1)}$ has $S(\bm{q}^{(t)}) = Z_{t-1}/Z_t$. A simple calculation shows that $J_t = J(\bm{q}^{(t)},\bm{u}) = (t-1) (\langle \bm{E} \rangle_{\bm{u}} - \langle \bm{E} \rangle_{\bm{q}^{(t)}})$. Substituting $S(\bm{q}^{(t)})$ and $J_t$ into \cref{eq: mean ep em} yields \cref{eq: mean ep for tilt}.

\bibliographystyle{apsrev4-2}

%

\end{document}